\documentclass[10pt,journal,compsoc]{IEEEtran}
\ifCLASSOPTIONcompsoc
  \usepackage[nocompress]{cite}
\else
  \usepackage{cite}
\fi
\ifCLASSINFOpdf
\else
\fi
\usepackage{algorithm}
\usepackage{algorithmic}
\usepackage{amsmath}
\usepackage{graphicx}
\usepackage[inkscapelatex=false]{svg}
\usepackage{amsfonts}
\usepackage{latexsym}
\usepackage{amsmath}
\usepackage{multirow}
\usepackage{subfig}
\usepackage[font=small,labelfont=bf]{caption}

\begin{document}
%
\title{Learning Service Selection Decision Making Behaviors During Scientific Workflow Development}
%
%
%
%

\author{Xihao~Xie,~
        Jia~Zhang,~\IEEEmembership{Senior Member,~IEEE},~
        Rahul~Ramachandran,~
        Tsengdar~J.~Lee,~
        Seungwon~Lee
\IEEEcompsocitemizethanks{
\IEEEcompsocthanksitem X. Xie, J. Zhang are with the Department of Computer Science, Southern Methodist
University, Dallas, TX, USA. 
E-mail: xihaox@smu.edu, jiazhang@smu.edu
\IEEEcompsocthanksitem R. Ramachandran is with the Marshall Space Flight Center, NASA, Huntsville, AL, USA. 
E-mail: rahul.ramachandran@nasa.gov
\IEEEcompsocthanksitem T. J. Lee is with the Science Mission Directorate, NASA Headquarters, Washington, DC, USA. 
E-mail: tsengdar.j.lee@nasa.gov
\IEEEcompsocthanksitem S. Lee is with the Jet Propulsion Laboratory, NASA, Pasadena, CA, USA. 
E-mail: seungwon.lee@jpl.nasa.gov
}
\thanks{Manuscript received June 20, 2023.}}

%
%

\markboth{IEEE Transactions on XX,~Vol.~XX, No.~XX, XX~202X}%
{Shell \MakeLowercase{\textit{et al.}}: Bare Demo of IEEEtran.cls for Computer Society Journals}
%




\IEEEtitleabstractindextext{%
\begin{abstract}
Increasingly, more software services have been published onto the Internet, making it a big challenge to recommend services in the process of a scientific workflow composition. In this paper, a novel context-aware approach is proposed to recommending next services in a workflow development process, through learning service representation and service selection decision making behaviors from workflow provenance. Inspired by natural language sentence generation, the composition process of a scientific workflow is formalized as a step-wise procedure within the context of the goal of workflow, and the problem of next service recommendation is mapped to next word prediction. Historical service dependencies are first extracted from scientific workflow provenance to build a knowledge graph. Service sequences are then generated based on diverse composition path generation strategies. Afterwards, the generated corpus of composition paths are leveraged to study previous decision making strategies. Such a trained goal-oriented next service prediction model will be used to recommend top K candidate services during workflow composition process. Extensive experiments on a real-word repository have demonstrated the effectiveness of this approach.

\end{abstract}

}

\maketitle

\IEEEdisplaynontitleabstractindextext

%
\IEEEpeerreviewmaketitle

\IEEEraisesectionheading{
\section{Introduction}
\label{sec:introduction}
}

%
%
%
%
\IEEEPARstart{I}{n} recent years, increasingly more software programs have been deployed and published onto the Internet as reusable web services, or so-called Application Programming Interfaces (APIs). Scientific researchers can thus leverage and compose existing services to build new data analytics experiments, so-called scientific workflow or workflow in short \cite{b1}. From a practical perspective, exploiting existing services instead of reinventing the wheel will lead to higher efficiency and productivity. However, our studies over the life science field \cite{b2}, which is one of the pioneer fields that adopt the concept of software service reuse, have revealed that the reusability rate of life science services in workflows remains rather low. Until 2018, less than 10\% of life science services published at biocatalog.org \cite{b3} were ever reused in scientific workflows, according to myExperiment.org \cite{b4}. This means that most of the published life science services have never been reused by anyone.


\begin{figure}[htbp]
\centerline{\includegraphics[width=0.45\textwidth]{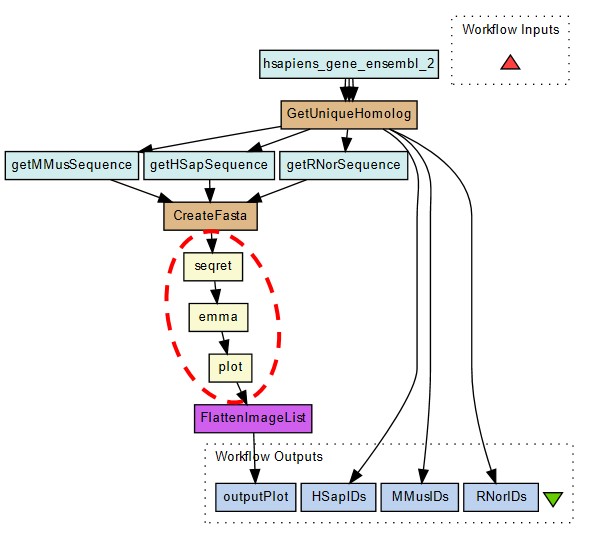}}
\caption{Motivating example workflow marked on \#1794 in myExperiment.org\protect\footnotemark}
\label{fig_motivating_example}
\vspace{-15pt}
\end{figure}
\footnotetext{https://www.myexperiment.org/workflows/1794.html}

What is wrong? Why scientists do not like to leverage others' work? Earlier studies believe one of the major obstacles making researchers unwilling to reuse existing services is the data shimming problem \cite{b5} \cite{b6}, meaning transforming the output data types of an upstream service to feed in the required inputs of a downstream service. For example, a NASA Climate Model Diagnostic Analyzer (CMDA) service, 2-D Variable Zonal Mean service that generates a graph of a 2-dimensional variable's zonal mean with time averaging, requires more than a dozen of input parameters \cite{b5}. Unless a user fully understands each parameter, both of its syntactic and semantic meaning and requirements, she may not feel comfortable to reuse the service.

To tackle the shimming problem in the context of  service recommendation, our earlier studies constructed a knowledge graph \cite{b4}\cite{b7} from all historical data of service invocations, i.e., service provenance, extracted from workflow repositories such as myExperiment.org. Various random walk-powered algorithms were developed to traverse the knowledge graph and suggest possible service candidates during workflow development \cite{b4}\cite{b7}.

In this paper, we move one step further, to argue that the selection of a software service is not only dependent on its direct upstream services, but also the context of all composed services in the current workflow under construction. Take the workflow \#1794 from myExperiment.org in Fig. \ref{fig_motivating_example} as an example. As highlighted in red oval, three services \textit{seqret}, \textit{emma}, and \textit{plot} are used sequentially to fetch a sequence set, do multiple alignments, and then plot the results. The adoption of the third service \textit{plot} aims to visualize the outcome from the two sequential services upstream in the workflow.

In order to take contextual information of workflow into consideration for more precise service recommendation in a recommend-as-you-go manner, this research presents a novel technique, inspired by Natural Language Processing (NLP), over service provenance knowledge graph. To be more specific, we formalize a process of workflow composition as a sequential step-wise process of service selection going towards the goal requirement of the workflow. We further formalize the problem of service recommendation as a problem of next service prediction, where services and composition paths of workflows are mapped to ``tokens" and sequential ``sentences" in the field of NLP, respectively. Our rationale is that, each path (i.e., a sequence of services) in a workflow reflects a scenario of data analytics experiment, which is analogous to a sentence in a conversation. In this way, our research goal turns into predicting and recommending the next suitable services that might be used for a user during the process of workflow composition.

The topic of next item prediction and recommendation has been studied extensively, and the literature has witnessed many successful applications in real-world fields such as e-commerce \cite{b9}, 
keyboard prediction \cite{b11} and sequential click prediction \cite{b12}. In general, two distinct categories of approaches exist to recommend next items in a sequential context. Traditional approaches are typically based on Markov chain (MC). For example, Rendle et al. \cite{b13} model sequential data by learning a personalized transition graph over underlying MC to predict and recommend items that a user might want to purchase. In recent years, a number of machine learning-based approaches have been proposed for next item recommendation \cite{b9} \cite{b14}. 

Our method falls into the second category, and applies machine learning techniques to learn service embedding based on their historical usages under various contexts. Given a workflow under construction, for the purpose of recommending next suitable services, we propose a two-stage approach to learning service representation and further predict next services: offline representation learning and online recommendation. At the offline stage, we employ a sequential modeling technique to learn representations of services by embedding them into a fix-sized vector space with a goal-oriented sequence modeling component. Without losing generality, we extend the traditional Long Short-Term Memory (LSTM) model into a goal-oriented context-aware LSTM (gLSTM), and incorporate it with an attention mechanism as a tailored sequence modeling component. Specifically, we first build a knowledge graph based on a workflow repository. According to our formalization of workflow composition, we propose two strategies, an \emph{intra-workflow} strategy and an \emph{inter-workflow} strategy, to extract and generate sequential composition paths each being analogous to a ``sentence" in NLP by traversing over the constructed knowledge graph. Afterwards, we feed the generated composition paths corpus and the goal requirements of workflows into the offline module of service representation learning. Such a learning process not only learns the embedding of services but also decision making strategies of service selection in the context of workflow goals. At the online stage, the learned model will predict the probabilities of services at run time in each step of workflow composition and produce top K candidate services in descending order. 


To the best of our knowledge, this is the first attempt to seamlessly exploit NLP, knowledge graph and the state-of-the-art sequential modeling techniques to facilitate service recommendation and workflow composition. Our extensive experiments over a real-world dataset have demonstrated the effectiveness of our approach. In summary, our contributions are three-fold:
\begin{itemize}
\item We formalize the process of workflow composition as a sequential process of service selection and further formalize the service recommendation problem as a problem of context-aware next service prediction.
\item We develop a goal-oriented sequential model, namely gLSTM for next service recommendation in a workflow development context.
\item We develop an approach to learning service representations and service selection decision making strategies from workflow provenance, which can guide recommend-as-you-go service selection during workflow development.
\end{itemize}

The remainder of this paper is organized as follows. Section 2 discusses related work. Section 3 states preliminary concepts and defines the research problem. Section 4 presents our framework of context-aware service recommendation in detail. Section 5 discusses and analyzes experimental results. Finally, Section 6 draws conclusions.

\vspace{-15pt}
\section{Related Work}
Our work is closely related to three categories of research in the literature: service recommendation, representation learning, and next item prediction.

\vspace{-15pt}
\subsection{Service Recommendation}
Service composition remains a fundamental research topic in the services computing community. Paik et al. \cite{b16} decompose service composition activities into four phases: planning, discovery, selection and execution. Service recommendation represents a core technology in the third phase.

Zhang et al. \cite{b7} model services, workflows and their relationships from historical usage data into a social network, to proactively recommend services in a workflow composition process. In their later work \cite{b4}, they develop an algorithm to extract units of work (UoW) from workflow provenance to recommend potentially chainable services. 
By leveraging NLP techniques, Xia et al. \cite{b20} propose a category-aware method to cluster and recommend services for automatic workflow composition. Shani et al. \cite{b8} formalize the problem of generative recommendation as a sequential optimization problem and apply Markov Decision Processes (MDPs) to solve the problem.

In the Business Process Management (BPM) community, a number of research work have explored methods to recommend reusable components for workflow composition. VisComplete \cite{b21} recommend components in a workflow as a path extension, by building graphs for workflows. Deng et al. \cite{b22} extract relation patterns between activity nodes from existing workflow repository to recommend extending activities. Zhang et al. \cite{b23} mine the upstream dependency patterns for workflow recommendation. Similarly, Smirnov et al. \cite{b24} specify action patterns using association rule mining to suggest additional actions in process modeling.

In contrast, our work formalize the process of workflow composition as a sequential step-wise process of service selection towards the goal requirement of the workflow. Subsequently, the problem of next service recommendation within a given context of workflow under construction is formalized as a problem of goal-oriented, context-aware next item recommendation.

\vspace{-10pt}
\subsection{Representation Learning}

In the services computing field, earlier work mainly focus on learning service representations based on natural language texts of service profiles. 
Li et al. \cite{b30} propose an approach to recommending services by analyzing semantic compatibility between user requirements and service descriptions. 
Zhong et al. \cite{b32} apply the Author-Topic Models 
to extract words as service description from mashup profiles. Zhang et al. \cite{b34} develop a tailored topic model to learn service representation for accurate visualization. Wang et al. \cite{b55} propose a method to recommend a group of services by learning embedding of entities based on truncated random walks in a knowledge graph.

Different from topic modeling-based methods, Service2vec \cite{b41} construct a service network 
and apply the Word2Vec modeling technique to extract contextual similarity between web services. 
Menzi et al. \cite{b45} represent services in a low-dimensional vector space based on a service knowledge graph to support downstream service recommendation applications. Xie et al. \cite{b35} leverage the Skip-Gram model \cite{b15} to learn representations of services based on service token sequences that are extracted from workflow provenance.

In contrast, our work differs from current literature of service representation and recommendation in three significant aspects. First, we believe that service usage context hides in paths in workflow provenance. Second, we formalize the problem of service recommendation as a problem of next service prediction in sequential context and have explored different sequence generation strategies. Third, complementary to existing service representation, we learn service representation and service selection decision making strategies from the context and sequential dependencies of services in workflows instead of profiles of services.

\vspace{-10pt}
\subsection{Next Item Prediction}
In our study, we employ sequence modeling techniques to model sequential composition behaviors and then recommend next service given a workflow construction context. From a technical perspective, existing sequence models can be divided into two categories: traditional models and neural network models.

Traditional sequence models are generally based on the Markov model (MM). 
For example, Rendle et al. \cite{b13} model users' sequential shopping behaviors by learning a transition graph based on MM for personalized recommendation. Zhang et al. \cite{b46} combine MM and content search techniques to model sequential web page clickstream data and recommend web pages. Based on variable-order Markov model (VMM), Garcin et al. \cite{b47} build a sequence context tree to recommend related news.

In recent years, neural network models such as Long Short-Term Memory (LSTM) \cite{b51} and Gated Recurrent Unit (GRU) have achieved great success in sequence modeling and have been widely used in the scenario of sequential recommendation. For example, Hidasi et al. \cite{b14} propose a GRU-based approach to take the sequential clickstream as input and predict the probability of the next item. Wu et al. \cite{b49} predict future behavior trajectories with an LSTM autoregressive model, by modeling the sequential evolution of items and changes in user preferences over time. It is worth mentioning that researchers have recently applied the attention mechanism to make sequential models more comprehensive. 
In the services computing field, Shi et al. \cite{b48} incorporate a functional attention mechanism and a contextual attention mechanism with the LSTM model to recommend services.

In contrast to existing work, this research develops gLSTM, a goal-oriented sequential modeling component, and leverages the attention mechanism to model the contribution scales of contextual services to finally recommend next services towards the goal requirement of workflow. 

\vspace{-15pt}
\section{Problem Formulation}
\subsection{Preliminary}
\newtheorem{definition}{Definition}
\begin{definition}[Service Repository]
A service repository is a collection of services formulated as $\mathcal{S} = \{s_{1}, s_{2}, ..., s_{|\mathcal{S}|}\}$, where $|\mathcal{S}|$ is the total number of services. Each service $s_{i} = \left<n_{i}\right>$ is unique from others with its textual name $n_{i}$.
\end{definition}
\begin{definition}[Workflow Repository]
A workflow repository is a set of workflows represented as $\mathcal{W} = \{w_{1}, w_{2}, ..., w_{|\mathcal{W}|}\}$, where $|\mathcal{W}|$ is the total number of workflows. A workflow $w_{i} \in \mathcal{W}$ is represented as a Directed Acyclic Graph (DAG) $G^{i} = \left<S^{i}, E^{i}, R_{i}\right>$ in terms of its topology. $S^{i} = \{s^{i}_{1}, s^{i}_{2}, ..., s^{i}_{|S^{i}|}\} \subseteq \mathcal{S}$ is a collection of service nodes composed in $w_{i}$. $E^{i} = \{e^{i}_{1}, e^{i}_{2}, ..., e^{i}_{|E^{i}|}\} \subseteq \mathcal{E}$ is a set of edges, where $e^{i}_{k} = \left<s^{i}_{u}, s^{i}_{v}, w_{i}\right> \in E^{i}$ is an edge, labeled with $w_{i}$, linking \emph{source} service node $s^{i}_{u}$ and \emph{sink} service node $s^{i}_{v}$ to indicate a directed dependency between them in $w_{i}$. $\mathcal{E}$ is the set of all edges between any two service nodes in $\mathcal{S}$. $R_{i}$ is the textually described goal requirement of $w_{i}$.
\end{definition}
\begin{definition}[Incremental Workflow Composition]
Following the definition of workflow, for a workflow $w_{i}$, at its creation step $t$, it is represented as a sub-graph $G^{i}_{t} = \left<S^{i}_{t}, E^{i}_{t}, R_{i}\right>$ of $G^{i}$, where $S^{i}_{t} \subset S^{i}$ is a set of services that have been selected at step $t$ and $E^{i}_{t} \subset E^{i}$ is a set of edges in $E^{i}$. At the next step $t + 1$, $w_{i}$ is incrementally composed by focusing on an \emph{anchor} service $s^{i}_{t} \in S^{i}_{t}$ and selecting a service $s^{i}_{t + 1} \in \mathcal{S} \setminus S^{i}_{t}$ as a sink service node of $s^{i}_{t}$, forming a new edge $e^{i}_{t + 1} = \left<s^{i}_{t}, s^{i}_{t + 1}, w_{i}\right> \in \mathcal{E} \setminus E^{i}_{t}$. Here, an \emph{anchor} service is the service to which the workflow composer wants to link a new service in terms of the expected topology. Thus, $S^{i}_{t + 1} = S^{i}_{t} \cup \{s^{i}_{t + 1}\}$, $E^{i}_{t + 1} = E^{i}_{t} \cup \{e^{i}_{t +1}\}$, and $G^{i}_{t + 1} = \left<S^{i}_{t + 1}, E^{i}_{t + 1}, R_{i}\right>$.
\end{definition}
\begin{definition}[Composition Path]
Over the workflow repository $\mathcal{W}$, a composition path is represented as a list of edges: $P = [e_{1}, e_{2}, ..., e_{|P|}]$, where $e_{k} \in \mathcal{E}$ is an edge in $\mathcal{E}$. For $\forall{e_{k}}$, its source service node is the sink node of $e_{k - 1}$. Thus, the starting and terminating service nodes of $P$ are the source service node of $e_{1}$ and the sink service node of $e_{|P|}$, respectively. Particularly, an \emph{anchor composition path} is defined as a path whose terminating service is the specified \emph{anchor} service. Otherwise, it is defined as a \emph{secondary composition path} of the anchor service.
\end{definition}
Note that, a collection of multiple composition paths that are generated following the definition might result in a fig-mental but working workflow which is nonexistent in $\mathcal{W}$. It enables our approach to make practical recommendation and in the meantime to increase the diversity of the recommended result. We will discuss the details in later sections.
\begin{definition}[Composition Context]
We assume that, while composing a workflow $w$, selecting a new service after an anchor service depends on not only the \emph{inner} sequential context of selected services along an anchor composition path $P$, but also the \emph{external} context of $P$. We define two kinds of composition contexts to be considered while composing a workflow, i.e., \emph{path-level composition context} and \emph{workflow-level composition context}, which are the list of services that have been selected in $P$ and other contextual information of $w$ that can be used for recommending next service after $P$, respectively.
\end{definition}
Consider that the recommended next service should not be a service which has been composed in $w$. Thus, in this study, for simplicity, we leverage the \emph{excluded services set} of $P$ in $w$ as the workflow-level composition context for next service prediction and recommendation. Specifically, for any pair of two composition paths, e.g., $P_{m}$ and $P_{n}$, let $S(P_{m}, P_{n})$ denote the intersection set of services between them. Given an anchor composition path $P_{m}$, the set of excluded services of it is represented as $S_{ex}(P_{m}) = \bigcup_{P_{n} \in \mathcal{P}^{S}}S(P_{m}, P_{n}) \subset \mathcal{S}$, where $\mathcal{P}^{S}$ is a set of secondary composition paths.

\vspace{-10pt}
\subsection{Problem Definition}
As mentioned earlier, we regard the process of composing a workflow as a step-wise process of service selection. Thereafter, the research problem of our study is described as: given a workflow $w' \notin \mathcal{W}$ which has not been completed yet at a specific time step $t$, our goal is to recommend a list of services that are suitable to be composed as a \emph{sink} service of the specified \emph{anchor} service at the next time step $t + 1$.

According to previous definitions, our research problem can be formalized as a problem of predicting a service $s$ to be selected at the next step $t + 1$ given an \emph{anchor} service $s'_{t}$ of a workflow $w' \notin \mathcal{W}$ that is under construction:
\begin{equation}
{p(s'_{t + 1}=s | G'_{t}, s'_{t}) = p(s | G'_{t}, s'_{t})}
\label{eq_1}
\end{equation}



\vspace{-15pt}
\section{Service Recommendation Methodology}
In order to solve the problem defined in Eq. \ref{eq_1}, we propose an approach of dividing it into composition path-based sub-problems, each being next service prediction given a sequential anchor composition path $P'_{t}$ and its excluded services set carrying partial context information of $G'_{t}$, and conquering them individually to finally recommend next services. In this section, we first depict the overview of our proposed framework. Then, we discuss each major component in the framework in detail, including offline knowledge graph construction, sequential composition path generation, sequence modeling and online service recommendation.

\begin{figure*}[htbp]
\centering
\includegraphics[width=0.8\textwidth]{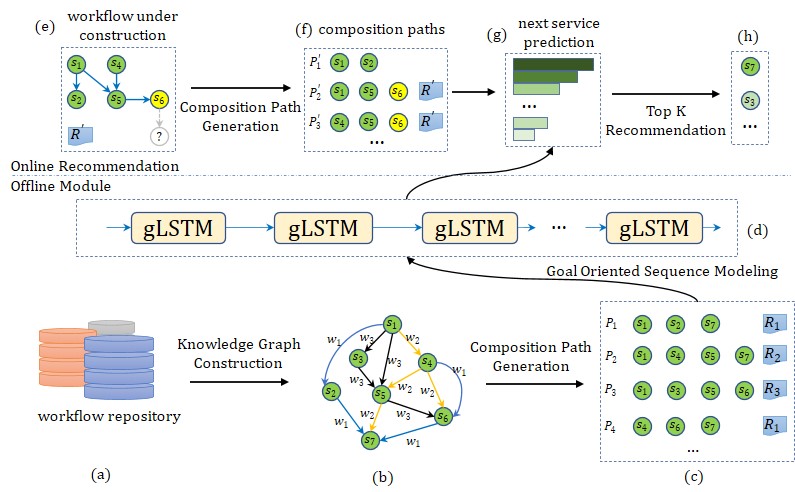}
\caption{Blueprint of proposed approach. (a) Workflow repository. (b) Constructed knowledge graph. (c) Composition paths generated in forms of service sequences. (d) Composition path-based service prediction model trained offline. (e) Real-time workflow under construction. (f) Composition paths generated for the ongoing workflow. (g) Predicted probabilities of potential services to be composed at the next step. (h) Recommended list of top K candidate services. Operations from (a) to (d) are conducted in the offline phase and (e)-(h) is the online recommendation phase.}
\label{fig_architecture}
\vspace{-15pt}
\end{figure*}

Fig. \ref{fig_architecture} presents the high-level blueprint of our methodology. Based on workflow provenance (a), we first construct a knowledge graph (b) by extracting service dependencies from workflows. Second, from the knowledge graph, we generate sequences of services (c) each denoting a composition path. Third, we train a composition path-based model (d) being able to predict the probabilities of services to be selected at the next step given an anchor composition path and goal requirement of workflow. As shown in Fig. \ref{fig_architecture}, the above steps are conducted in the offline module. 
In the online recommendation module, given an anchor service (colored in yellow) in a workflow under construction (e), we generate its anchor composition paths and secondary composition paths (f) and then predict the probabilities of next services that have not been composed with the trained model (d). Next, we rank the potential services according to their predicted probabilities in descending order (g), and finally recommend top K of them (h) to users at real time. 

\vspace{-10pt}
\subsection{Knowledge Graph Construction}
In this subsection, we introduce the details of constructing the service knowledge graph (SKG) from workflow repository. Similar to our earlier method described in detail in \cite{b7}, SKG is a directed graph (digraph) carrying historical service invocation dependencies extracted from workflow repository, and it is defined as:
\begin{equation}
SKG = \left<\mathcal{N}, \mathcal{R}\right>
\label{eq_2}
\end{equation}
where each service $s \in \mathcal{N}$ is regarded as an entity node, and $\mathcal{R} = \left\{r^{w_{i}}_{u,v}\right\}$ is a set of relationships between entities. $r^{w_{i}}_{u, v} = \left<s_{u}, s_{v}, w_{i}\right>$ refers to a relationship that $s_{u}$ is an upstream service node of $s_{v}$ in workflow $w_{i}$. We can regard the relationship $r^{w_{i}}_{u,v}$ as an edge starting from $s_{u}$ and ending at $s_{v}$ with label $w_{i}$. Note that there might be multiple edges between two service nodes in the knowledge graph with different labels, meaning that such a service invocation dependency happens in multiple workflows. For example, as shown in Fig. \ref{fig_motivating_example}, service \emph{emma} is used as a downstream service of \emph{seqret} in workflow \#1794. Meanwhile, the dependency relationship between them exists in workflow \#2226\footnote{https://www.myexperiment.org/workflows/2226.html} as well.

Naturally, following the definition in preliminary, SKG can be built from workflow repository by leveraging the set of services (i.e., $\mathcal{S}$) and the set of edges (i.e., $\mathcal{E}$) as the collections of entities and relationships, respectively. Therefore, SKG can be represented as:
\begin{equation}
SKG = \left<\mathcal{S}, \mathcal{E}\right>
\label{eq_3}
\end{equation}

\vspace{-15pt}
\subsection{Sequential Composition Path Generation}
How to extract and generate composition paths from workflow repository is a key step in our approach. Different generation strategies may result in different recommendations. 
In this subsection, we explore two different ways, each of which might indicate a specific composition behavior, to generate sequential composition paths from the constructed knowledge graph SKG introduced. The two generation strategies are: intra-workflow generation and inter-workflow generation. Fig. \ref{fig_tangible_workflow} is a portion extracted from the constructed from SKG that motivates and will be used to explain 
the two generation strategies.

\vspace{-8pt}
\subsubsection{Intra-workflow Generation}
In this method, we consider individual workflows. The composition paths generated with the intra-workflow restriction make it possible to model the composition behaviors of published workflows. Specifically, we regard the process of composing a workflow $w_{i} \in \mathcal{W}$ as a process of generating sequential composition paths, each of which is a sequence of services traversing along edges with label $w_{i}$ in SKG from a starting service without a source service in $w_{i}$. 
Formally, let $P^{i}_{u, m} = [e^{i}_{1}, e^{i}_{2}, ..., e^{i}_{|P^{i}_{u, m}|}] = s_{u} \xrightarrow[]{w_{i}} ... \xrightarrow[]{w_{i}} s_{k} \xrightarrow[]{w_{i}} ... \xrightarrow[]{w_{i}} s_{v}$ denote a generated composition path, starting from $s_{u} \in S^{i}$ and going to next unvisited neighbor services along edges with label $w_{i}$, until meeting a terminal service $s_{v} \in S^{i}$ without a successor service along label $w_{i}$. $m \in \left[1, M\right]$ and $M$ is the total number of composition paths that can be generated starting from $s_{u}$.

\begin{figure*}[htbp]
\centering
\includegraphics[width=0.8\textwidth]{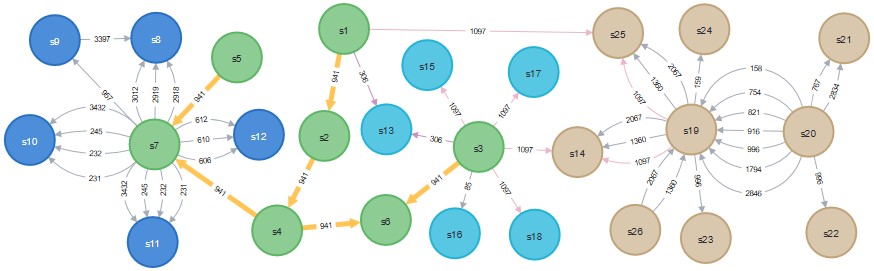}
\caption{Portion of SKG motivating two service sequence generation methods. Edges are labeled with correponding id numbers of workflows. The nodes in green are services in workflow \#941\protect\footnotemark and the dependencies between them are colored in orange with label ``941." Nodes in other colors are services invoked by other workflows, which appear to be downstream nodes of the services in workflow \#941 in SKG.}
\label{fig_tangible_workflow}
\vspace{-15pt}
\end{figure*}

Take the workflow \#941 from myExperiment.org for a simple example. For illustration purpose, Fig. \ref{fig_tangible_workflow} uses $s_{1}$, $s_{2}$, ..., $s_{7}$ to stand for the services nodes of \emph{String\_Constant}, \emph{getPeak\_input}, \emph{String\_Constant1}, \emph{getPeak}, \emph{String\_Constant0}, \emph{XPath\_From\_Text0}, and \emph{XPath\_From\_Text}, respectively. In the corresponding workflow, the nodes $s_{1}$, $s_{3}$, $s_{5}$ are starting service entities and nodes $s_{6}$, $s_{7}$ are terminal service entities that can be traversed along the label ``941." Applying the intra-workflow generation strategy, we can firstly generate four sequential composition paths, two of which starting from $s_{1}$: $P^{941}_{1, 1} = s_{1}\xrightarrow[]{941}s_{2}\xrightarrow[]{941}s_{4}\xrightarrow[]{941}s_{6}$, $P^{941}_{1, 2} = s_{1}\xrightarrow[]{941}s_{2}\xrightarrow[]{941}s_{4}\xrightarrow[]{941}s_{7}$, $P^{941}_{3, 1} = s_{3}\xrightarrow[]{941}s_{6}$ and $P^{941}_{5, 1} = s_{5}\xrightarrow[]{941}s_{7}$. Afterwards, four more composition paths can be generated starting from intermediate services: $P^{941}_{2, 1} = s_{2} \xrightarrow[]{941} s_{4} \xrightarrow[]{941} s_{6}$, $P^{941}_{2, 2} = s_{2} \xrightarrow[]{941} s_{4} \xrightarrow[]{941} s_{7}$, $P^{941}_{4, 1} = s_{4} \xrightarrow[]{941} s_{6}$, $P^{941}_{4, 2} = s_{4} \xrightarrow[]{941} s_{7}$.

It can be proven that, in workflow $w_{i}$, given an anchor composition path $P_{m}$ terminating at service $s_{v}$, for $\forall{s_{j}} \in S_{ex}(P_{m})$, it may not be a sink service of $s_{v}$. If not, there must be a cyclic path $s_{j} \xrightarrow[]{w_{i}} ... \xrightarrow[]{w_{i}} s_{v} \xrightarrow[]{w_{i}} s_{j}$, which is not allowed to form a workflow. Thus, using $S_{ex}(P_{m})$ as the external context of $P_{m}$ to indicate a set of services that should be excluded when predicting next services is meaningful and practical.

\footnotetext{https://www.myexperiment.org/workflows/941.html}

\vspace{-8pt}
\subsubsection{Inter-workflow Generation}
The intra-workflow generation strategy is a restricted method to generate sequential composition paths. It focuses on modeling how an existing workflow is composed. Thus, when recommending next services for a workflow under construction, the predicted result will make the workflow tend to be similar to an existing workflow. However, the exploratory feature of scientific data analytics naturally demands diversity and unprecedented logic comprising new service chaining. Thereafter, in the inter-workflow generation method, we loosen the restriction by traversing SKG without the restriction of going along with specific labels of existing workflows. Inspired by \textsc{DeepWalk} \cite{b37}, which uses random walk to generate sequences of nodes 
in a graph, we leverage a generalized variant of random walk, i.e., probabilistic walk, to generate sequential inter-workflow composition paths based on the constructed knowledge graph SKG.

In contrast to the intra-workflow generation strategy, the inter-workflow generation strategy considers the generation of next service rooted at $s_{u}$ as a stochastic process, with random variables $S^{1}_{u}, S^{2}_{u}, ..., S^{l}_{u}$ where $S^{l+1}_{u}$ is a service generated with probabilities from the neighbors of service $s_{l}$. Note that, such a service $S^{l+1}_{u}$ existing in the sequence of $\left[ S^{1}_{u}, S^{2}_{u}, ..., S^{l}_{u} \right]$ is not allowed, meaning that the generated sequence is acyclic. On the one hand, restarting from a service does not show any improvement to our experimental results. On the other hand, it is generally meaningless to invoke a previously invoked service in the same workflow. Specifically, for a service $s_{u}\in{\mathcal{S}}$, let $N(s_{u}) \subseteq{\mathcal{S}}$ denote the set of directed neighbor services in the whole SKG, we model the probability $p_{u,v}$ that service $s_{v}$ can be generated after $s_{u}$ with the commonly used softmax transformation:
\vspace{-5pt}
\begin{equation}
\vspace{-5pt}
p_{u,v} = p(s_{v}|s_{u})=\frac{\exp(\frac{o_{u,v}}{o_{u}})}{\sum_{s_{n}\in{N(s_{u})}}\exp(\frac{o_{u,n}}{o_{u}})}
\label{eq_puv}
\end{equation}
where $s_{v}\in{N(s_{u})}$, $o_{u,v}$ is the number of occurrence of relationships between $s_{u}$ and $s_{v}$ in the whole SKG and $o_{u} = \sum_{n\in{N(s_{u})}}o_{u,n}$. Note that we can degenerate the probabilistic walk method into the random walk method, by sampling all directed neighbor services of $s_{u}$ equally as a sink service.

For example, in Fig. \ref{fig_tangible_workflow}, the dependency relationship between $s_{7}$ and $s_{10}$ occurs four times in four different workflows whose id numbers are ``3432," ``245," ``232" and ``231," respectively. However, for $s_{9}$, it appears after $s_{7}$ only once in workflow ``957." Therefore, according to Eq. \ref{eq_puv}, the probability that $s_{10}$ can be generated as a downstream service of $s_{7}$ is greater than that of $s_{9}$.

In this method, every edge $e_{k}$ in a generated path $P = [e_{1}, e_{2}, ..., e_{|P|}]$ is a tangible dependency relationship in SKG, which means that $P$ is a working composition path of a workflow that might be composed by others in the future, even though $P$ is not observed in the workflow repository. In a scenario of service recommendation, holding it out from composition paths is defective and perfunctory. In the meantime, the generated inter-workflow composition paths grant recommendation higher diversity. For a simple example, in Fig. \ref{fig_tangible_workflow}, $P = [e_{1}, e_{2}]$ is a possible composition path, where $e_{1} = \left<s_{5}, s_{7}, w_{941}\right>$ and $e_{2} = \left<s_{7}, s_{11}, w_{232}\right>$. In the workflow repository, $s_{5}$, $s_{7}$ and $s_{11}$ are never composed in the same workflow. However, $s_{7}$ and $s_{11}$ are observed as sink services of $s_{5}$ and $s_{7}$, respectively in different workflows, indicating that they might be composed in a future workflow.

Two factors may impact the effectiveness of the inter-workflow generation strategy. One factor is the length of a visiting path, i.e., $l$, which determines when to stop while traversing the SKG. The other factor is the number of walks starting at each service, i.e., $\tau$. In practice, without specifying $\tau$, some linkages may not be traversed. As a result, the generated sequences may not cover all tangible dependency relationships, tampering the ability of trained representations to predict next suitable services. As \cite{b37} suggested, although it is not strictly required to do this, it is a common practice to tune the two factors to speed up the offline training stage. In the later section of experiments, we will discuss in detail the effects of $l$ and $\tau$.

Note that, in this study, for a composition path $P$ generated with the inter-workflow strategy, to ensure that a recommended service could not be a service which has been composed, we simply use the set of services in itself as the excluded services set (i.e., workflow-level context).

\vspace{-5pt}
\subsubsection{Discussions}
In this subsection, we discuss more details of sequential composition path generation in the above two aspects. First, as shown in Fig. \ref{fig_architecture}, sequence generation is conducted in both offline learning stage and online recommendation stage. During the offline training phase, we apply either intra-workflow or inter-workflow strategy over the SKG to generate composition paths in forms of service sequences. However, during the online phase, only the workflow under construction is used as the input to trigger service recommendation. Thus, for real-time online recommendation, we apply the intra-workflow generation strategy to extract composition paths.

Second, for both aforementioned strategies, no matter which one we apply, there might appear duplicate composition paths. The reason is that some services and their dependency relationships in a workflow $w_{i}$ might exist in another workflow $w_{j}$, in a form of service chain. For example, in Fig. \ref{fig_tangible_workflow}, applying the intra-workflow generation strategy, composition path $s_{26} \xrightarrow[]{} s_{19} \xrightarrow[]{} s_{25}$ can be generated twice through two workflows, i.e., with labels ``1360" and ``2067."

\begin{figure}
\centerline{\includegraphics[width=0.45\textwidth]{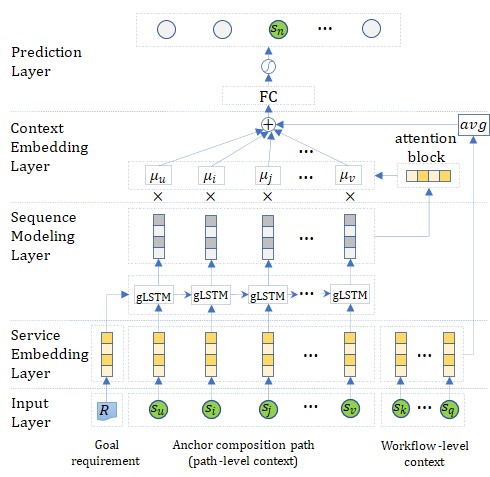}}
\caption{Architecture of composition path-based service prediction.}
\label{fig_sequence_modeling}
\vspace{-10pt}
\end{figure}

As for the inter-workflow generation strategy, three factors may result in duplicates: the length of a path, the frequencies of the dependency relationships in the path,  and the predefined parameter $\tau$. A shorter path with more frequent dependency relationships tends to be generated as a sequence with higher probability in case of a higher $\tau$. For example, in Fig. \ref{fig_tangible_workflow}, let $s_{20}$, $s_{19}$ and $s_{14}$ stand for services \emph{seqret}, \emph{emma} and \emph{emma\_NJ}, respectively. The frequency of the dependency relationship between $s_{20}$ and $s_{19}$ is much higher than those between $s_{20}$ and other services. So does that of the relationship between $s_{19}$ and $s_{14}$. Therefore, the sequence of composition path $s_{20} \xrightarrow[]{} s_{19} \xrightarrow[]{} s_{14}$, whose length is three, can be generated multiple times.

Similar to insurmountable duplicates in a natural language corpus \cite{b36}, duplicates in the corpus of sequential composition paths generated from the knowledge graph might impact the effectiveness of our approach. It is a \emph{feature-but-not-bug} problem. We will evaluate the effect of duplicates in the later section of experiments.

\vspace{-12pt}
\subsection{Context-based Prediction}
As stated previously, to solve the problem of \emph{which services we should select at the next step given an anchor service in a workflow under construction}, we first solve the sub-problem of \emph{which service we should select at the next step given an anchor composition path containing a sequence of selected services}. As shown in Fig. \ref{fig_architecture}, given a corpus of generated composition paths in forms of service sequences, we train our prediction model with sequence modeling techniques which will be leveraged for online service recommendation. 

Fig. \ref{fig_sequence_modeling} illustrates the architecture of our context-based service prediction model, in a layered framework. Given the goal requirement, an anchor composition path (i.e., path-level context) and its excluded services set, we first represent the goal requirement and services as fix-sized real-value vectors. Then, to capture both the dependencies between services in the anchor composition path and the goal requirement, we develop a goal-oriented LSDM, i.e., gLSTM extending the LSTM \cite{b56} to partially model the sequential behavior of workflow composition towards the goal requirement. Next, considering that different services in a sequence of services make different contribution scales to representing the composition path and further predicting next services, we apply an attention mechanism to represent the sequential context of selected services into a vector. At the same time, as the workflow-level context of the anchor composition path, the embeddings of excluded services are averaged and integrated to represent the whole context in a fully connected layer. Afterwards, for services that have not been selected in the sequential context, we model the likelihoods of them to be selected at the next step with a fully connected layer, and normalize the corresponding output with softmax transformation. In summary, the offline training phase of context-based service prediction consists of four consecutive steps: input representation, goal-oriented sequence modeling, attention-based context embedding, and context-aware next service prediction.

\vspace{-18pt}
\subsubsection{Input Representation}
\vspace{-5pt}
In Fig. \ref{fig_sequence_modeling}, the sequence modeling component takes a generated composition path and the goal requirement of the workflow as input. Each service in the composition path is embedded into a fix-size vector space, by representing it as a $d$-dimension real-valued vector. Formally, let $\mathbf{W}^{S} \in \mathbb{R}^{d \times \left|\mathcal{S}\right|}$ be the embedding matrix of services. Thus, a service $s_{i}$ can be represented as vector $\mathbf{W}^{S}_{:,i}$. The embedding matrix of services, i.e., $\mathbf{W}^{S}$, is randomly initialized and will be learned.

To represent the goal requirement that is textually described in natural language, we employ the Doc2Vec \cite{b57} model to transform it into a fix-sized vector $\mathbf{V}^{R} \in \mathbb{R}^{d \times 1}$. Each textual description of workflow goal requirement is regarded as a document. The corpus of all documents is fed into the Doc2Vec model to receive a vectorized representation of each goal requirement. Specially, we adopt the distributed memory model of paragraph vectors (PV-DM) as it performs better than the distributed bag of words of paragraph vector (PV-DBOW) \cite{b57}.


\vspace{-20pt}
\subsubsection{Goal-Oriented Sequence Modeling}
\begin{figure}
\centerline{\includegraphics[width=0.4\textwidth]{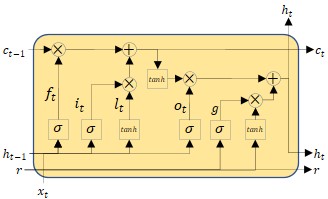}}
\caption{Inner structure of the gLSTM memory block.}
\label{fig_glstm}
\vspace{-12pt}
\end{figure}

\vspace{-5pt}
We extend the LSTM \cite{b56} into goal-oriented LSTM (gLSTM) to model the sequential dependencies between services in a composition path towards the goal requirement. Fig. \ref{fig_glstm} presents the inner structure of a gLSTM memory block. $r$ is the vectorized goal requirement, i.e., $\mathbf{V}^{R}$. $x_{t}$ is the input at time $t$. $c_{t}$ is the global state sharing different cell outputs throughout the whole gLSTM networks. $h_{t}$ outputs the transformed result of $x_{t}$ at time $t$ and $r$. $f_{t}$, $i_{t}$ and $o_{t}$ are the forget, input and output gates, respectively. $g$ is the goal gate to transform the goal requirement so that $h_{t}$ carries the information of the goal requirement. 

As shown in Fig. \ref{fig_sequence_modeling}, for an edge $e = \left<s_{i}, s_{j}, w\right>$ in a composition path $P$, the source service $s_{i}$ and the sink service $s_{j}$ appearing at time $t - 1$ and $t$, respectively, the representation carrying dependency information of $s_{j}$ and the goal requirement is calculated by feeding their embeddings into gLSTM to conduct a series of transformations as below:
\begin{equation}
\begin{aligned}
&f_{t}(s_{j}) = \sigma(\mathbf{W}_{xf} \cdot \mathbf{W}^{S}_{:,j} + \mathbf{W}_{hf} \cdot h_{t - 1}(s_{i}) + b_{f}) \\
&i_{t}(s_{j}) = \sigma(\mathbf{W}_{xi} \cdot \mathbf{W}^{S}_{:,j} + \mathbf{W}_{hi} \cdot h_{t - 1}(s_{i}) + b_{i}) \\
&l_{t}(s_{j}) = \tanh(\mathbf{W}_{xl} \cdot \mathbf{W}^{S}_{:,j} + \mathbf{W}_{hl} \cdot h_{t - 1}(s_{i}) + b_{l}) \\
&c_{t}(s_{j}) = f_{t}(s_{j}) c_{t - 1}(s_{i}) + i_{t}(s_{j}) l_{t}(s_{j}) \\
&o_{t}(s_{j}) = \sigma(\mathbf{W}_{xo} \cdot \mathbf{W}^{S}_{:,j} + \mathbf{W}_{ho} \cdot h_{t - 1}(s_{i}) + b_{o}) \\
&g(s_{j}) = \sigma(\mathbf{W}_{g} \cdot \mathbf{V}^{R} + b_{g}) \\
&z(s_{j}) = \tanh(\mathbf{W}_{z} \cdot \mathbf{V}^{R} + b_{z}) \\
&h_{t}(s_{j}) = o_{t}(s_{j}) \tanh(c_{t}(s_{j})) + g(s_{j}) z(s_{j})
\label{eq_10}
\end{aligned}
\end{equation}
where $\sigma$ and $\tanh$ are the sigmoid function $\sigma(x) = 1 / (1 + \exp(-x))$ and the hyperbolic tangent activation function $\tanh(x) = (\exp(x) - \exp(-x)) / (\exp(x) + \exp(-x))$ respectively, $\mathbf{W}_{x*} = \{\mathbf{W}_{xf},\mathbf{W}_{xi},\mathbf{W}_{xl},\mathbf{W}_{xo}\}$, $\mathbf{W}_{h*} = \{\mathbf{W}_{hf},\mathbf{W}_{hi},\mathbf{W}_{hl},\mathbf{W}_{ho}\}$ and $\mathbf{W}_{gz} = \{\mathbf{W}_{g}, \mathbf{W}_{z}\}$ are the weights of the corresponding gates of forget, input, output and goal, and $b_{*} = \{b_{f},b_{i},b_{l},b_{o}, b_{g}, b_{z}\}$ is bias. They are also randomly initialized and will be learned. In this way, the final representation of $s_{j}$ in a sequence at position $t$ carrying information of both the sequential context and the goal requirement is formed 
as $h(s_{j}) = h_{t}(s_{j})$.

\vspace{-8pt}
\subsubsection{Attention-based Context Embedding}
In order to model different contribution scales of selected services in an anchor composition path, we apply an attention mechanism to model their importance in representing the whole sequential context. As shown in Fig. \ref{fig_sequence_modeling}, for a service $s_{i}$ in an anchor composition path, its attention weight is assigned with the softmax transformation as:
\vspace{-5pt}
\begin{equation}
\mu(s_{i}) = \frac{\exp(a(s_{i}))}{\sum_{s_{k} \in S^{P}}\exp(a(s_{k}))}
\label{eq_11}
\end{equation}
\begin{equation}
a(s_{i}) = \mathbf{A} \cdot h(s_{i})
\label{eq_12}
\vspace{-5pt}
\end{equation}
where $S^{P}$ is the set of services in the anchor composition path $P$, and $\mathbf{A} \in \mathbb{R}^{1 \times d}$ is a $d$-dimension vector in the attention block in Fig. \ref{fig_sequence_modeling}. It is randomly initialized and will be learned as well.

In the meantime, the workflow-level context of $P$ is represented as a $d$-dimension vector by averaging the embeddings of the excluded services of $P$. As a result, the whole context containing the anchor composition path $P$ in a form of sequential context of selected services and the workflow-level context of excluded services is represented as a $d$-dimension vector:
\begin{equation}
\vspace{-15pt}
v(P) = \sum_{s_{i} \in S^{P}} \mu(s_{i}) \times h(s_{i}) + \frac{\sum_{s_{k} \in S_{ex}(P)}\mathbf{W}^{S}_{:,k}}{|S_{ex}(P)|}
\label{eq_13}
\end{equation}

\subsubsection{Context-Aware Service Prediction}
As shown in Fig. \ref{fig_sequence_modeling}, to predict which service should be selected next given a sequential composition path $P$, we feed its representation $v(P)$ into a fully connected layer and normalize the output as probabilities of services to be selected at the next step. Let $\mathbf{W}^{F} \in \mathbb{R}^{\left|\mathcal{S}\right| \times d}$ denote the weight matrix of the fully connected layer. The likelihood of service $s_{n}$ to be selected next is modeled with the softmax transformation as:
\vspace{-5pt}
\begin{equation}
p(s = s_{n} | P) = \frac{\exp(r(s_{n}))}{\sum_{s_{k} \in \mathcal{S}}\exp(r(s_{k}))}
\label{eq_14}
\end{equation}
\vspace{-5pt}
\begin{equation}
r(s_{n}) = \mathbf{W}^{F}_{n,:} \cdot v(P)
\label{eq_15}
\vspace{-5pt}
\end{equation}
Here, $r(s_{n})$ can be considered as a scoring function qualifying the relevance of a
service $s_{n}$ with respect to the given composition path $P$. Similarly, $\mathbf{W}^{F}$ is randomly initialized and will be learned.

\vspace{-5pt}
\subsubsection{Parameters Learning}
In the sequence modeling component, $\mathbf{\Theta} = \mathbf{W}_{x*} \cup \mathbf{W}_{h*} \cup \mathbf{W}_{gz} \cup b_{*} \cup \{\mathbf{W}^{S}, \mathbf{W}^{F}, \mathbf{A}\}$ is a set of parameters to be learned over a training set of composition paths $\mathcal{P} = \{P_{1}, P_{2}, ..., P_{|\mathcal{P}|}\}$ that is generated from the SKG. The joint probability distribution can be obtained as:
\begin{equation}
p_{\mathbf{\Theta}}(\mathcal{P}) \propto \prod_{P_{i} \in \mathcal{P}} p(s = s_{j} | \widetilde{P_{i}})
\label{eq_16}
\vspace{-10pt}
\end{equation}
where $\widetilde{P_{i}}$ is the sub-sequence of $P_{i}$ by eliminating the last edge $e_{|P_{i}|}$ from $P_{i}$ and $s_{j}$ is the terminating service of $P_{i}$, i.e., the sink service of $e_{|P_{i}|}$. Thus, parameters in $\mathbf{\Theta}$ can be learned by maximizing the following cumulative objective:
\begin{equation}
\underset{\mathbf{\Theta}}{\arg\max}\log \prod_{P_{i} \in \mathcal{P}} p(s = s_{j} | \widetilde{P_{i}})
\label{eq_17}
\end{equation}
However, according to Eq. \ref{eq_14}, optimizing the objective in Eq. \ref{eq_17} is intractable since each evaluation of the softmax function has to traverse all services, which is of high computation cost. To learn $\mathbf{\Theta}$ efficiently, we employ the idea of negative sampling \cite{b53} to approximate the conditional probability in Eq. \ref{eq_16} as:
\begin{equation}
p_{\mathbf{\Theta}}(s_{j}, N^{j} | \widetilde{P}) = \prod_{s_{k} \in \{s_{j}\} \cup N^{j}} p_{\mathbf{\Theta}}(s_{k} | \widetilde{P})
\label{eq_18}
\end{equation}
\vspace{-5pt}
\begin{equation}
p_{\mathbf{\Theta}}(s_{k} | \widetilde{P}) = 
\begin{cases}
\sigma(r_{\mathbf{\Theta}}(s_{k})) & s_{k} \in \{s_{j}\} \\
1 - \sigma(r_{\mathbf{\Theta}}(s_{k})) & s_{k} \in N^{j}
\end{cases}
\label{eq_19}
\end{equation}
in which $s_{j}$ is the ground truth service, $N^{j} \subseteq \mathcal{S}\setminus\{s_{j}\}\setminus S_{ex}(\widetilde{P})$ is a set of negative services randomly sampled, and $r_{\mathbf{\Theta}}(x)$ corresponds to the scoring function defined in Eq. \ref{eq_15}. Here, $\sigma(x)$ is the sigmoid function. It can be seen as the probability of service $s_{k}$ being labeled as the ground truth service with a score of $r_{\mathbf{\Theta}}(s_{k})$ given $\widetilde{P}$. Thus, the probability of it being labeled as a negative service is $1 - \sigma(r_{\mathbf{\Theta}}(s_{k}))$. In this way, maximizing the conditional probability in Eq. \ref{eq_18} means maximizing the probability of a service being a positive sample and minimizing the probability of it being a negative sample at the same time. Consequently, the optimization objective function in Eq. \ref{eq_17} becomes:
\begin{equation}
\begin{aligned}
\mathcal{L} &= \log \prod_{P_{i} \in \mathcal{P}} \prod_{s_{k} \in \{s_{j}\} \cup N^{j}} p_{\mathbf{\Theta}}(s_{k} | \widetilde{P_{i}}) = \sum_{P_{i} \in \mathcal{P}} \mathcal{L}_{i} \\
& = \sum_{P_{i} \in \mathcal{P}} \{ \log[\sigma(r_{\mathbf{\Theta}}(s_{j}))] + \sum_{s_{k} \in N^{j}} \log[1 - \sigma(r_{\mathbf{\Theta}}(s_{k}))] \}
\label{eq_20}
\end{aligned}
\end{equation}
Then, given a composition path $P_{i}$ in the training set $\mathcal{P}$, we adopt stochastic gradient ascent to update parameters in $\{\mathbf{W}^{S}, \mathbf{W}^{F}, \mathbf{A}\}$ as follows:
\begin{equation}
\theta' \gets \theta + \eta \cdot \frac{\partial \mathcal{L}_{i}}{\partial \theta}, \forall{\theta} \in \{\mathbf{W}^{S}, \mathbf{W}^{F}, \mathbf{A}\}
\label{eq_21}
\end{equation}
where $\eta$ is the learning rate. For parameter $\vartheta \in \mathbf{W}_{x*} \cup \mathbf{W}_{h*} \cup \mathbf{W}_{gz} \cup b_{*}$ of the gLSTM model, we update it with Back-propagation Through Time (BPTT) \cite{b54}. To be more specific, the update rule of $\vartheta$ at time $t$ is:
\vspace{-3pt}
\begin{equation}
\begin{aligned}
\frac{\partial \mathcal{L}_{i}}{\partial \vartheta_{t}} &= (\delta_{t} + \frac{\partial \mathcal{L}_{i}}{\partial \vartheta_{t + 1}}) \cdot \frac{\partial h(s_{t})}{\partial \vartheta_{t}} \\
\vartheta_{t}' &\gets \vartheta_{t} + \eta \cdot \frac{\partial \mathcal{L}_{i}}{\partial \vartheta_{t}}
\end{aligned}
\label{eq_22}
\end{equation}
\vspace{-3pt}
in which $s_{t}$ is the service appearing at time $t$ of composition path $P_{i}$, $h(s_{t})$ is the corresponding output from LSTM networks, and $\delta_{t}$ is the update that is passed down from time steps after $t$.

Algorithm 1 describes the procedure of parameters learning offline. In the beginning, we randomly initialize the set of parameters $\mathbf{\Theta}$ to be learned (line 1). Then, we construct the knowledge graph SKG from the workflow repository $\mathcal{W}$ (line 2). Meanwhile, the set of all services $\mathcal{S}$ is extracted from $\mathcal{W}$. Next, we generate a set of composition paths $\mathcal{P}$ from the constructed knowledge graph SKG with the specified generation strategy (line 3). In each iteration, for each anchor composition path $P_{i} \in \mathcal{P}$, we generate a training instance of $P_{i}$, containing the ground truth service $s_{j}$, a set of negative services randomly sampled from $\mathcal{S}$, the sub-sequence $\widetilde{P_{i}}$ of $P_{i}$, and the excluded services set of $P_{i}$ (line 11). Then, according to Eq. \ref{eq_21}, we update parameters $\mathbf{A}$ and their corresponding weights in $\mathbf{W}^{F}$ for ground truth service $s_{j}$ and all negative services in $N^{j}$ (line 14). Similarly, we update the embeddings of excluded services of $\widetilde{P_{i}}$ (line 17). Afterwards, for every source service $s_{u}$, appearing at time $t$ ($t$ ranges from $|\widetilde{P_{i}}|$ to 1), of edges in $\widetilde{P_{i}}$, we update the parameters of the gLSTM networks according to Eq. \ref{eq_22} (line 21). In the meantime, according to Eq. \ref{eq_21}, we update the embedding of $s_{u}$, i.e., $\mathbf{W}^{S}_{:,u}$ (line 23). We stop updating parameters in case the accumulative objective which is defined in Eq. \ref{eq_20} converges or the number of iteration $c$ is greater than the specified maximum iteration number $L$. In the end, we obtain the learned parameters set $\mathbf{\Theta}$ to support online service recommendation.

\vspace{-10pt}
\subsection{Online Service Recommendation}
As discussed above, the trained sequence model enables us to know \emph{which service we should compose given a sequential anchor composition path}. This further enables us to predict \emph{which services we should select at the next step after the specified anchor service while composing a workflow}. In this subsection, we introduce the online prediction and recommendation module of our approach.

Given an anchor service $s'_{t}$ of a DAG $G'_{t}$ of workflow $w' \notin \mathcal{W}$ that is still under construction at creation step $t$, with the aforementioned intra-workflow generation strategy, we first extract a set of anchor composition paths $\mathcal{P}'_{t} = \{P'_{t, 1}, P'_{t, 2}, ..., P'_{t, |\mathcal{P}'_{t}|}\}$ from $G'_{t}$, each of which terminates at $s'_{t}$. Then, for a service $s_{n} \in \mathcal{S}$, we calculate its probability to be selected next as the sink service of the terminate service of $P'_{t, k}$, i.e., $p(s = s_{n} | P'_{t, k})$, according to Eq. \ref{eq_14}. After that, the probability of $s_{n}$ to be selected at the next step $t + 1$ that is defined in Eq. \ref{eq_1} is calculated as:
\begin{equation}
p_{\mathbf{\Theta}}(s'_{t + 1} = s_{n} | G'_{t}, s'_{t}) = \frac{\sum_{P'_{t, k} \in \mathcal{P}'_{t}}p_{\mathbf{\Theta}}(s'_{t + 1} = s_{n} | P'_{t, k})}{|\mathcal{P}'_{t}|}
\label{eq_online_rec}
\end{equation}
where $p_{\mathbf{\Theta}}(s'_{t + 1} = s_{n} | P'_{t, k})$ is the conditional probability defined in Eq. \ref{eq_14}. Finally, we rank services in $\mathcal{S}$ according to their calculated probabilities in descending order and recommend top K of them as the list potential services that are suitable to be composed at the next step after $s'_{t}$.

Note that, according to Eq. \ref{eq_14}, $p(s'_{t + 1} = s_{n} | P'_{t, k}) \in (0, 1)$ and $p(s'_{t + 1} | P'_{t, k}) = \sum_{s_{n} \in \mathcal{S}}p(s'_{t + 1} = s_{n} | P'_{t, k}) = 1$. Thereafter, for any service ${s_{n}} \in \mathcal{S}$, its conditional probability $p_{\mathbf{\Theta}}(s'_{t + 1} = s_{n} | G'_{t}, s'_{t}) \in (0, 1)$ is guaranteed. In the meantime, for all services in $\mathcal{S}$, the sum of their conditional probabilities, i.e., $p_{\mathbf{\Theta}}(s'_{t + 1} | G'_{t}, s'_{t}) = \sum_{s_{n} \in \mathcal{S}}p_{\mathbf{\Theta}}(s'_{t + 1} = s_{n} | G'_{t}, s'_{t})$, is guaranteed to be 1 as well.

\begin{algorithm}[H]
\caption{Parameters Learning Offline}
\begin{algorithmic}[1]
\renewcommand{\algorithmicrequire}{\textbf{Input:}}
\renewcommand{\algorithmicensure}{\textbf{Output:}}
\REQUIRE workflows $\mathcal{W}$, dimension size $d$, composition path generation strategy $\mbox{G}$, learning rate $\eta$, and maximum iteration number $L$
\ENSURE set of parameters $\mathbf{\Theta}$
\STATE Initialize the set of parameters $\mathbf{\Theta}$ randomly
\STATE $\left<\mathcal{S}, SKG\right> \gets BuildKnowledgeGraph(\mathcal{W})$
\STATE $\mathcal{P} \gets GenerateCompositionPaths(SKG, \mbox{G})$
\STATE $c \gets 0$
\WHILE{not converged \rm{and} $c \leq L$}
    \STATE $c \gets c + 1$
    \FOR{each composition path $P_{i} \in \mathcal{P}$}
    {
      \IF{$P_{i}$ is a secondary composition path}
      \STATE continue
      \ENDIF
      \STATE $\left<s_{j}, N^{j}, \widetilde{P_{i}}, S_{ex}(\widetilde{P_{i}})\right> \gets TraniningSample(P_{i}, \mathcal{S})$
      \FOR{each service $s_{k} \in \{s_{j}\} \cup N^{j} $}
      {
        \FOR{each parameter $\theta \in \{\mathbf{W}^{F}_{k,:}, \mathbf{A}\}$}
        {
          \STATE $\theta \gets \theta + \eta \cdot \frac{\partial \mathcal{L}_{i}}{\partial \theta}$
        }
        \ENDFOR
        \FOR{each excluded service $s_{q} \in S_{ex}(\widetilde{P_{i}})$}
        \STATE $\mathbf{W}^{S}_{:, q} \gets \mathbf{W}^{S}_{:, q} + \eta \cdot \frac{\partial \mathcal{L}_{i}}{\partial \mathbf{W}^{S}_{:, q}}$
        \ENDFOR
        \FOR{each service $s_{u}$ appearing at time $t$ in $\widetilde{P_{i}}$}
        {
          \FOR{each $\vartheta \in \mathbf{W}_{x*} \cup \mathbf{W}_{h*} \cup \mathbf{W}_{gz} \cup b_{*}$}
          {
            \STATE $\vartheta_{t} \gets \vartheta_{t} + \eta \cdot \frac{\partial \mathcal{L}_{i}}{\partial \vartheta_{t}}$
          }
          \ENDFOR
          \STATE $\mathbf{W}^{S}_{:,u} \gets \mathbf{W}^{S}_{:,u} + \eta \cdot \frac{\partial \mathcal{L}_{i}}{\partial \mathbf{W}^{S}_{:,u}}$
        }
        \ENDFOR
      }
      \ENDFOR
    }
    \ENDFOR
\ENDWHILE
\RETURN $\mathbf{\Theta}$
\label{alg_offline_learning}
\end{algorithmic}
\end{algorithm}

\vspace{-15pt}
\section{Experiments}
We have designed and conducted a collection of experiments to evaluate our proposed approach. In this section, we first provide an overview of the dataset used, then explain the evaluation metrics adopted, and then present the experimental results and analyses.

\subsection{Dataset}
Started from 2007, myExperiment.org has become the largest service-oriented scientific workflow repository in the world. It has been used as a testbed by many researchers \cite{b4}, \cite{b5}, \cite{b7} in the services computing community. Since we construct the knowledge graph SKG across workflow boundaries, we target on the Taverna-generated workflows, i.e., workflows generated following Taverna syntax.

In the target testbed, every workflow is maintained as an XML file and every service in a workflow is defined as a \textit{processor}. A dependency relationship between two services is defined as a \textit{link} or \textit{datalink} whose two ends are \textit{source} and \textit{sink}, respectively. We examined all Taverna workflows published on myExperiment.org up to December 2021, with a total of 2,910 workflows and 8,780 services. Table \ref{tab1} summarizes the statistical information over all workflows in the dataset.

\begin{table}[htbp]
\caption{Statistics about myExperiment Dataset}
\vspace{-10pt}
\label{tab1}
\begin{center}
\begin{tabular}{|r|r|c|}
\hline
\multicolumn{2}{|r|}{Total \# of workflows} & 2,910 \\\hline
\multicolumn{2}{|r|}{Total \# of services} & 8,780 \\\hline
\multicolumn{2}{|r|}{Avg. \# of services per workflow} & 12.07 \\\hline
\multicolumn{2}{|r|}{Avg. \# of sink services per service} & 1.60 \\\hline
\multicolumn{2}{|r|}{Avg. \# of intra-workflow composition paths per workflow} & 15.90 \\\hline
\multicolumn{2}{|r|}{Avg. length of all composition paths} & 9.60 \\\hline
\end{tabular}
\end{center}
\end{table}
\vspace{-5pt}

\vspace{-15pt}
\subsection{Experimental Setup}
We randomly selected 80\% of the workflows as the training set to build the SKG and the remaining 20\% of workflows as the test data. For the intra-workflow generation strategy, all services in a workflow in the training set were used as anchor services to generate composition paths. As for the inter-workflow generation strategy, all services in the SKG were used as starting services to generate composition paths with specified $l$ and $\tau$. The generated sequential composition paths with length smaller than 2 were removed from training.

In the offline learning phase, we set the learning rate $\eta$, the dimension size $d$ and the maximum iteration number $L$ to be 0.001, 128 and 20, respectively. In the online recommendation phase, for each workflow in the testing set, we used every service of it as an anchor service to generate composition paths with the intra-workflow generation strategy and all sink services of the anchor service in the workflow as the ground truth to evaluate the accuracy of the recommended top $K$ services. Besides, we also evaluated the diversity of recommendation results by using all sink services of an anchor service in the SKG as the ground truth. Considering that our approach is unable to predict a service that is never seen in training set, for each testing workflow, we held out its composition paths containing any of such services.

For textual descriptions of goal requirements of workflows, we performed a series of pre-processing operations including tokenization, stop-words removal and lemmatization with the NLTK\footnote{http://www.nltk.org/} library.

\subsection{Evaluation Metrics}
As discussed previously, composition paths generated with different strategies over SKG model workflow composition behavior differently. 
Thus, different corpus of composition paths result in different recommendation performance. In this subsection, we introduce the accuracy metrics we adopted, followed by the diversity metric.

\subsubsection{Accuracy Metrics}
To evaluate the accuracy of our recommendation framework, we employed \textit{Recall@K} and \textit{MRR} as the evaluation metrics. They measure the recall of the top $K$ ranked services and the mean reciprocal rank of the ground truth service in the recommendation result over all testing cases,  respectively. For each workflow $w_{i}$ in the testing set, they can be calculated as follows:
\begin{equation}
Recall@K = \frac{1}{|S^{i}|} \times \sum_{s^{i}_{k} \in S^{i}} \frac{|R(s^{i}_{k}) \cap G(s^{i}_{k}) |}{|G(s^{i}_{k})|}
\end{equation}
\begin{equation}
MRR = \frac{1}{|S^{i}|} \times \sum_{s^{i}_{k} \in S^{i}} \frac{1}{rank(R(s^{i}_{k}), G(s^{i}_{k}))}
\end{equation}
where $R(s^{i}_{k})$ is the recommended result of anchor service $s^{i}_{k}$, $G(s^{i}_{k})$ is the corresponding ground truth and $rank(R(s^{i}_{k}), G(s^{i}_{k}))$ is the rank of the first service in $R(s^{i}_{k})$ hitting the ground truth. For both metrics, the higher the better. For metric $Recall@K$, we reported the results with $K \in \{3, 5, 10, 20\}$.

Note that, precision and F-measure, i.e., $Precision@K$ and $F1@K$ respectively, are commonly used evaluation metrics for recommendation methods. In our study, we hold that recall and MRR are better metrics than precision and F-measure to gauge the performance between different strategies. Recall that our goal is to find the most suitable services following a specific anchor service. It means that we care more about how many sink services we can fetch, instead of how many recommended services are sink services. The higher recall means higher possibility to help a user increase the efficiency of service composition. Besides, for the rank of the ground truth service, we hope it could be as low as possible. The higher MRR could encourage users to reuse the recommended services much more. For example, if a service exists in the ground truth, e.g., $s_{i}$ is followed by two services, say $s_{j}$ and $s_{k}$, which rank at 1st and 4th of the recommended list, respectively. The precision is only 40\%. However, it is acceptable for the user in real practice to compose one of $s_{j}$ and $s_{k}$ as a sink service of $s_{i}$, especially in a scenario that there are more than hundreds of services in the whole repository, which makes it tedious and time-consuming to find a suitable service at the next step.

\begin{table*}[htbp]
\caption{Overall Recommendation Results}
\vspace{-8pt}
\label{tab_overall_result}
\begin{center}
\begin{tabular}{c|c|c|c c c c|c c c c|c}
\hline
\multirow{2}{*}{Methods} & \multirow{2}{*}{Strategies} & {Duplicates} & \multicolumn{4}{c|}{Recall@K} & \multicolumn{4}{c|}{Diversity@K} & \multirow{2}{*}{MRR} \\ \cline{4-11}
& & Removed & K = 3 & K = 5 & K = 10 & K = 20 & K = 3 & K = 5 & K = 10 & K = 20 \\
\hline
\multirow{4}{*}{Ours} & Intra-workflow & No & \textbf{0.8825} & \textbf{0.9220} & \underline{0.9329} & \underline{0.9399} & 0.7375 & 0.7917 & 0.8255 & 0.8525 & \textbf{0.8084} \\
 & Inter-workflow & No & 0.7444 & 0.8062 & 0.8580 & 0.8831 & 0.7408 & 0.7994 & 0.8526 & 0.8811 & 0.6755 \\
 & Intra-workflow & Yes & \underline{0.8750} & 0.8989 & 0.9138 & 0.9244 & \underline{0.8137} & \underline{0.8583} & \underline{0.8975} & \underline{0.9209} & \underline{0.7881} \\
 & Inter-workflow & Yes & 0.8590 & \underline{0.9206} & \textbf{0.9619} & \textbf{0.9855} & \textbf{0.8484} & \textbf{0.9110} & \textbf{0.9603} & \textbf{0.9836} & 0.7584 \\
\hline
SG-RW & Random Walk & Yes & 0.6635 & 0.7291 & 0.7870 & 0.8252 & 0.7126 & 0.7580 & 0.8206 & 0.8485 & 0.7114 \\
SG-PW & Probabilistic Walk & Yes & 0.7249 & 0.7863 & 0.8416 & 0.8755 & 0.7282 & 0.7651 & 0.8196 & 0.8507 & 0.7321 \\
\hline
\end{tabular}
\end{center}
\vspace{-12pt}
\end{table*}

\subsubsection{Diversity Metric}
When selecting next service to an anchor service, it is not necessary to restrict a newly composed workflow to be similar to a specific historical workflow. On the contrary, recommending diversely working services at each composition step indicates more possibility to create working workflows and increase service reusability. We believe the recommended list of next services with higher diversity is better for workflow composers. For a service $s$, we hold that any sink service of it in historical workflows could be next service while composing a new workflow with $s$ as an anchor service. In our study, we propose a method to measure the diversity of the recommended list of next services given a workflow $w_{i}$ as:
\begin{equation}
Diversity@K = \frac{1}{|S^{i}|} \times \sum_{s^{i}_{k} \in S^{i}} \frac{\left|R(s^{i}_{k}) \cap SS(s^{i}_{k})\right|}{\left|SS(s^{i}_{k})\right|}
\end{equation}
where $SS(s^{i}_{k})$ is the set of all sink services of $s^{i}_{k}$ in the SKG. The higher the value of $Diversity@K$ is, the more the diversity of the recommended result is, and the more diverse the newly composed workflow would be. In our study, we reported the results of $Diversity@K$ with $K \in \{3, 5, 10, 20\}$.


\vspace{-10pt}
\subsection{Baseline Methods}
In our experiments, we used two methods as baselines for comparison:
\begin{itemize}
\item \textbf{SG-RW} \cite{b55}: In this method, an undirected knowledge graph is built for nodes of Mashups and APIs, which are regarded as workflows and services respectively in our study. Based on node sequences that are extracted from the knowledge graph with random walk, the Skip-gram model \cite{b15} is leveraged to learn node representations. Services are recommended according to the similarity scores between them and the target Mashup.
\item \textbf{SG-PW} \cite{b35}: Our earlier method was designed to recommend next services for workflow composition. It extracts sequential services as tokens of sentences and leverages the Skip-gram model \cite{b15} to learn service representations. Finally, it ranks and recommends next services according to service similarities in the vector space.
\end{itemize}
For SG-RW, considering that a workflow is a directed acyclick graph in this paper, we tailered the undirected knowledge graph as a directed graph. The types of services and workflows in our dataset were regarded as the categories of Mashups and APIs in SG-RW. We set the walk length, the number of walks per node, and the dimension size for SG-RW to be 40, 6 and 32, respectively. In \cite{b55}, the effect of duplicate node sequences were not investigated. In our experiments, we reported the recommendation results of SG-RW for duplicates being removed.

For SG-PW, we employed its probabilistic-walk (PW) generation strategy to extract service sequences. The path length, the number of paths per service and the dimension size in SG-PW were set to be 5, 10 and 50, respectively. According to their suggestions, we removed duplicate sequences from the training corpus to learn service representations.

\vspace{-10pt}
\subsection{Performance Evaluation}
Table \ref{tab_overall_result} reports the evaluation results by different recommendation performance metrics described above. For each metric, the best result is highlighted in bold and the second best is underlined. We designed three research questions to help assess our approach and analyze experimental results in different views: (1) to study which composition path generation strategy should be adopted; (2) to examine the effect of the generated duplicate composition paths; and (3) to investigate the effectiveness of our approach.

\emph{RQ1: Which composition path generation strategy is the best?}
According to the experimental results, it is hard to demonstrate which one is the best. Different generation strategies perform differently in terms of different evaluation metrics.

In terms of accuracy metrics, composition paths generated with the intra-workflow strategy performs better than that with the inter-workflow strategy when the recommendation list size $K$ is lower than 10. However, as $K$ becomes higher, the best of inter-workflow strategy gets better than that of the intra-workflow strategy (i.e., 0.9619 vs 0.9329 and 0.9855 vs 0.9399 for $K = 10$ and $K = 20$, respectively). It means that the rank of the ground truth is lower with lower $K$ for the intra-workflow strategy. In terms of MRR, we observed the same results. The best value using the intra-workflow strategy (i.e., 0.8084) is higher than that of the inter-workflow strategy (i.e., 0.7584). Thus, the intra-workflow generation strategy is a better option in case a smaller size of the recommendation list is expected.

In terms of diversity metric, the experimental results demonstrate that composition paths generated with the inter-workflow strategy shows its capability to recommend services with higher diversity. As explained earlier, the intra-workflow strategy models the composition patterns from existing workflows. However, the inter-workflow strategy tends to generate composition paths unprecedented. The diversity gap between both strategies highly depends on the similarities between the training workflows and the testing workflows.


To better understand our findings, let $R(D, D') = \frac{|D \cap D'|}{|D|}$ denote the ratio of shared identical composition paths in $D$, where $D$ and $D'$ are two sets of composition paths. A lower $R(D, D')$ indicates a higher similarity of $D'$ to $D$. Then, let $D_{t}$ be the set of composition paths coming from testing workflows, $D_{intra}$ and $D_{inter}$ be the sets of composition paths generated from the SKG with the intra-workflow and the inter-workflow strategies, respectively. In our study, $R(D_{t}, D_{intra})$ and $R(D_{t}, D_{inter})$ are 0.2659 and 0.1846, respectively. Reversely, $R(D_{intra}, D_{t})$ and $R(D_{inter}, D_{t})$ are 0.0498 and 0.0097 respectively. Thus, the generated composition paths with the inter-workflow strategy over the SKG is less similar to the composition paths generated from the testing workflows, making the recommended result more diverse.

In summary, different generation strategies might result in different recommendation results under various recommendation scenarios. In practice, it is encouraged to clarify the objective of the recommendation scenario to make proper decision of composition path generation strategy. Table \ref{tab_overall_result} also reveals that, no matter which strategy was adopted, as $K$ increases from 3 to 20, the overall mean recall ranges from 0.8402 to 0.9332. That is to say, almost 90\% of ground truth services were retrieved. In terms of diversity metric, its overall mean value ranges from 0.7851 to 0.9095. For an anchor service with 10 sink services in the SKG, approximately 80\% of them could be received as next services. In terms of MRR, its overall mean value is around 0.75, meaning that the ground truth service is ranked at 1st or 2nd approximately. The experimental results have demonstrated the effectiveness of our approach to recommending next services for workflow composition. We published a demo on Github showing how our proposed approach supports next service recommendation in workflow composition \footnote{https://github.com/DSI-SMU/WorkflowHelperDemo}.

\emph{RQ2: Should duplicates from the generated sequential composition paths be removed?} In NLP field, it has been discovered that different NLP models are affected by duplicates in different ways \cite{b36}. In this project, no matter which strategy is employed, there might be duplicate composition paths generated. Our approach to solving the sub-problem of predicting the next service given a sequential composition path can be analogized predicting the next ``word" given a sequence of words, where a service and a sequential sentence can be analogized as a word and a composition path, respectively. Thus, it is necessary to investigate the effect of duplicate composition paths in next service recommendation.

As shown in Table \ref{tab_overall_result}, for the inter-workflow strategy, holding duplicates out from the corpus makes better recommendation results in terms of all evaluation metrics. On the one hand, as we investigated in RQ1, the corpus of composition paths generated with the inter-workflow strategy tends to be less similar to that generated from testing workflows. Therefore, removing duplicates can prevent the corpus from being less similar to testing corpus, which makes a better recommendation result in terms of accuracy metrics. On the other hand, according to Eq. \eqref{eq_puv}, a higher probability to generate $s_{v}$ as a sink service of $s_{u}$ holds a higher probability to generate duplicate composition paths containing an edge which starts from $s_{u}$ and ends at $s_{v}$. In other words, remaining duplicates makes the trained model more likely to fit such edges which exist more frequently than others. Consequently, the trained model tends to recommend services which exist more frequently as sink services of the specific anchor service, resulting in lower diversity. Hence, it is encouraged to remove duplicates when generating training corpus of composition paths with the inter-workflow strategy.

However, for the intra-workflow strategy, removing duplicates increases recommendation diversity but conversely performs worse. Unlike the inter-workflow strategy, the intra-workflow strategy tends to generate composition paths being identical with that generated from testing workflows. Therefore, remaining duplicates could increase their similarity, making the trained model fit better to the testing corpus and resulting in higher accuracy but lower diversity.

Overall, duplicate composition paths generated from the SKG affect the recommendation result differently when different generation strategies employed. We suggest to investigate the impact of duplicate sequential composition paths in advance, when applying sequence modeling techniques to learn service representations in next service recommendation scenarios.

\emph{RQ3: How does our approach compare to baselines?} According to the experimental results shown in Table \ref{tab_overall_result}, our approach outperforms both SG-RW and SG-PW in all evaluation metrics.

Our method performs better than SG-RW maybe because of three reasons. First, the order of services in a composition path is an important factor that should be take into account for representation learning and recommendation. In the Skip-gram model, the order of services is ignored, which might result in recommending an upstream service as the next service for a given anchor service. Second, sequences generated with random walk cannot capture the weight information of service dependencies, which is valuable to generate tangible sequential composition paths. Third, we model the contribution scales of services in a composition path with an attention mechanism, which plays an important role in context representation in our approach.

As for SG-PW, as illustrated in \cite{b35}, a scoring function is defined to prevent it from recommending upstream services, making it performs better than SG-RW. However, it is not competitive to our approach. First, it uses only the information of a given anchor service for service recommendation. In our approach, we also take into account the information of composition paths all of which end at the anchor service. Second, in our approach, when representing a composition path, the contribution scales of services in it are also modeled with an attention mechanism, which makes our approach more expressive and adaptive.

In summary, our approach is effective in the scenario of next service recommendation for workflow composition. 

\begin{figure*} [t!hbp]
	\centering
    \subfloat[\label{fig_sensitivity_recall}]{
    \includegraphics[width=0.32\textwidth]{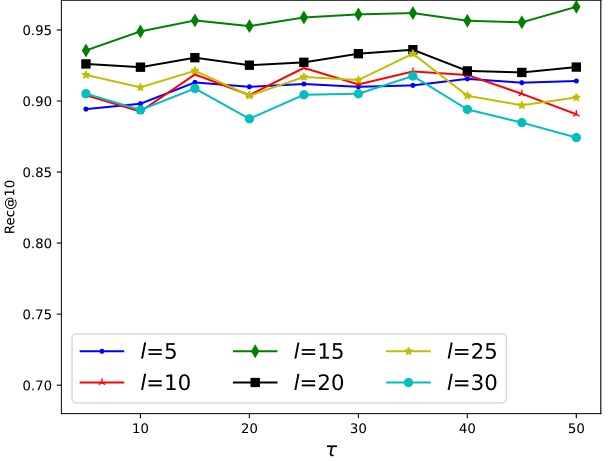}
    }
    \subfloat[\label{fig_sensitivity_diversity}]{
    \includegraphics[width=0.32\textwidth]{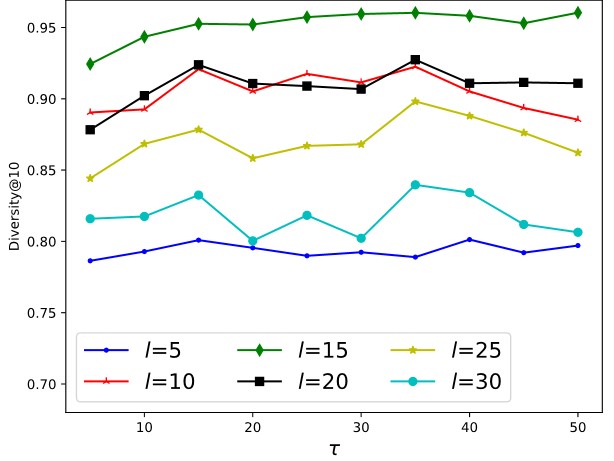}
    }
    \subfloat[\label{fig_sensitivity_mrr}]{
    \includegraphics[width=0.32\textwidth]{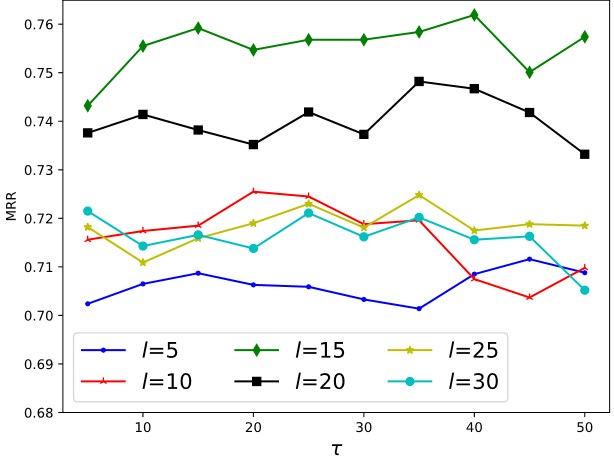}
    }
	\caption{The effect of $l$ and $\tau$ with top-10 recommendations}
	\label{fig_sensitivity} 
\end{figure*}

\vspace{-10pt}
\subsection{Parameter Sensitivity}
We designed experiments to evaluate how changes to $l$ and $\tau$ will effect the performance of the inter-workflow generation strategy. In such experiments, according to the lessons we learnt from RQ2, duplicate composition paths were removed. Figs. \ref{fig_sensitivity_recall}, \ref{fig_sensitivity_diversity} and \ref{fig_sensitivity_mrr} present the effects of $l$ and $\tau$ to recommendation performance with the inter-workflow generation strategy in terms of $Recall@10$, $Diversity@10$ and $MRR$, respectively. 

In terms of all evaluation metrics, our approach performs the best when $l = 15$. As $l$ increases from 5 to 15, the recommendation performance gets better. However, as $l$ increases from 15 to 30, the performance becomes worse. As shown in Table \ref{tab1}, the mean length of all composition paths generated with the intra-workflow strategy is 9.60, and the mean number of sink services over all services in the SKG is 1.60. Approximately, with the inter-workflow strategy, generating composition paths with length of $1.60 * 9.60 \approx 15$ can cover most composition paths generated with the intra-workflow strategy. A lower $l$ results in lower coverage over all linkages between services, which reduces the capability of discovering suitable services after a composition path ending at a specific anchor service. A higher $l$ makes it more likely to generate sequential composition paths that do not exist in testing workflows.

For the parameter $\tau$, as it increases from 5 to 50, the recommendation performance fluctuates slightly. Increasing $\tau$ does not necessarily result in significant improvements. It is hard to demonstrate the optimal $\tau$, however, we suggest to investigate its effects to recommendation performance.

In general, our experiments have demonstrated that $l$ and $\tau$ are two parameters that should be investigated in advance, when applying the inter-workflow strategy to generate composition paths during the offline training phase. To determine when to stop while traversing the SKG, we suggest to get a glimpse at the length of all composition paths that can be generated with the intra-workflow strategy and the out-degree (i.e., the number of sink services) of service nodes in the constructed knowledge graph.

\vspace{-5pt}
\section{Conclusions and Future Work}
In this paper, we have formalized workflow composition process as a goal-driven, context-aware sequential service generation process. A goal-oriented LSTM model (i.e., gLSTM) is developed to learn service representations and service selection decision making strategies, associated with information of path-level composition context and workflow-level composition context. The resulted service embeddings are used to support incremental workflow composition at run-time.

In the future, we plan to extend our research in the following two directions. First, we plan to take into account more composition context, such as users' profile information, to further enable personalized workflow recommendation. Second, we plan to seamlessly integrate sequential service invocation dependency with hierarchical graph structure to further enhance recommendation quality.

\vspace{-10pt}
\ifCLASSOPTIONcompsoc
  \section*{Acknowledgments}
\else
  \section*{Acknowledgment}
\fi
Our work is partially supported by National Aeronautics and Space Administration under grants 80NSSC21K0576,  80NSSC21K0253, and 80NSSC22K0144.

\ifCLASSOPTIONcaptionsoff
  \newpage
\fi



%

%

\begin{IEEEbiography}[{\includegraphics[width=1in,height=1.2in,clip,keepaspectratio]{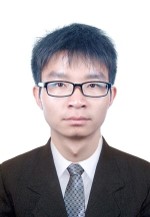}}]{Xihao Xie} received the B.S. degree in computer science from the University of Science and Technology of China and the M.S. degree in software engineering from University of Chinese Academy of Science, China. He is currently working toward the Ph.D. degree in the Department of Computer Science, Southern Methodist University, TX, USA. His research interests include service computing, recommender system, data science and data engineering.
\end{IEEEbiography}

\begin{IEEEbiography}[{\includegraphics[width=1in,height=1.2in,clip,keepaspectratio]{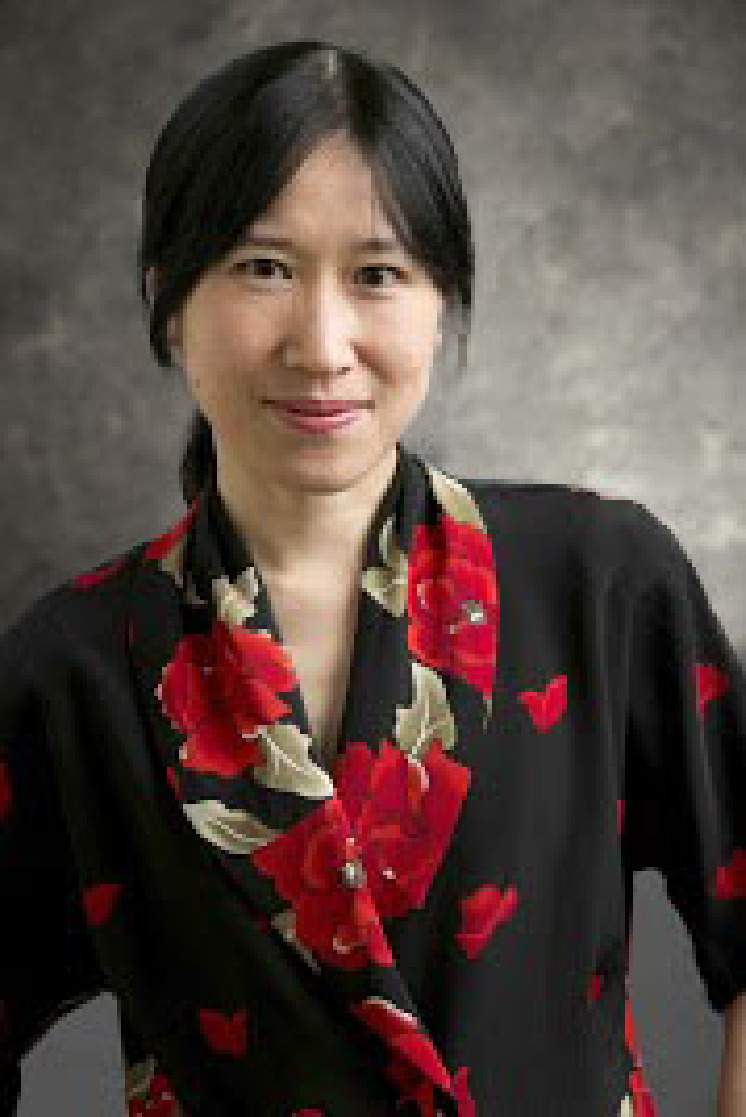}}]{Jia Zhang} received her M.S. and B.S. degree in computer science from Nanjing University, China and her Ph.D. degree in computer science from the University of Illinois at Chicago. She is currently Cruse C. and Marjorie F. Calahan Centennial Chair in Engineering, Professor at the Department of Computer Science, Southern Methodist University. Her recent research interests center on data science infrastructure, with a focus on scientific workflows, software discovery, and knowledge graph. She has co-authored one textbook titled ``Service Computing" and has published more than 170 referred journal paers, book chapters, and conference papers. She is currently an associate editor of the IEEE Transactions on Services Computing (TSC). She is a senior member of the IEEE.
\end{IEEEbiography}

\begin{IEEEbiography}[{\includegraphics[width=1in,height=1.2in,clip,keepaspectratio]{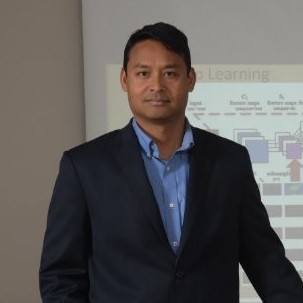}}]{Rahul Ramachandran} received a B.E. in Mechanical Engineering from Jamia Millia Islamia, India, an M.S. in South Dakota Mines, and an M.S. in Atmospheric Science, an M.S. in Computer Science, and a Ph.D. in Atmospheric Science from the University of Alabama in Huntsville. He is currently a Project Manager and Senior Research Scientist at National Aeronautics and Space Administration. His research focuses on Earth Science Informatics. He has 42 peer-reviewed publications including two book chapters and over 100 other scientific publications including workshop reports. He serves as the Deputy Editor for the Earth Science Informatics Journal (Springer). He received the Presidential Early Career Award for Scientists and Engineers (PECASE) award in 2009.
\end{IEEEbiography}

\begin{IEEEbiography}[{\includegraphics[width=1in,height=1.2in,clip,keepaspectratio]{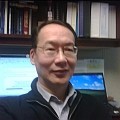}}]{Tsengdar Lee} received his M.S. degree in civil engineering and his Ph.D. in Atmospheric Science from Colorado State University. He is currently the Program Manager for the R\&A Weather Focus Area, the High-End Computing Program, and for NASA’s Data for Operation and Assessment. He Served as NASA Chief Technology Officer for Information Technology (CTO-IT) between 2011 and 2012. He set up the IT-Labs at NASA and invested in cloud computing and big data projects. His research focuses on the integration of weather and ancillary geographical information data into weather models to produce reliable forecasts. His research pioneered the modeling of land surface hydrology's impact on weather forecasting. He also serves as co-chair for the Interagency Weather Research Coordination Committee (IWRCC).
\end{IEEEbiography}

\begin{IEEEbiography}[{\includegraphics[width=1in,height=1.2in,clip,keepaspectratio]{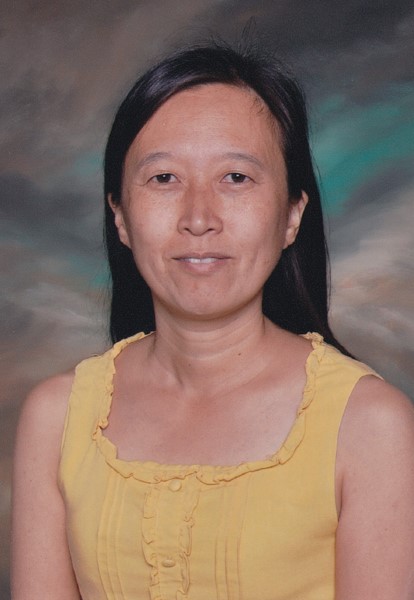}}]{Seungwon Lee} received her Ph.D. in Physics from Ohio State University. She is currently the technical group supervisor of the Science Data Modeling and Computing Group in Jet Propulsion Laboratory, California Institute of Technology. She has been leading research and technology development tasks for Earth science observation and modeling data analysis, science mission data processing system development, scientific data analysis, observation planning and operation, materials modeling and simulation, scheduling problems, spectral retrieval, cometary nucleus thermal modeling, cometary gas dynamics, remote sensing simulation, and flight-project formulation and proposal development.
\end{IEEEbiography}





\end{document}